\documentclass[mypaper,8pt,twoside]{CoAst}
\usepackage{epsf,graphicx,fancyhdr}
\renewcommand{\chaptermark}[1]{\markboth{#1}{}}

\newcommand{\ngc}[1]{NGC~{#1}}                              

\newcommand{\kopf}{\small\itshape Comm. in Asteroseismology\\ Vol. 144, 2003}
\newcommand{\Authors}[1]{\begin{center}\normalsize\bf\sf #1 \end{center}}

\renewcommand{\author}[1]{\begin{center}\normalsize\bf\sf #1 \end{center}}
\newcommand{\Address}[1]{\begin{center}\small\sf #1 \end{center}}

\renewenvironment{abstract}{\section*{Abstract}\normalsize\sf}{}
\newenvironment{introduction}{\section*{Introduction}\normalsize\sf}{}
\newenvironment{spheroidal}{\section*{Dwarf Spheroidal Galaxies}\normalsize\sf}{}
\newenvironment{irregular}{\section*{Dwarf Irregular Galaxies}\normalsize\sf}{}
\newenvironment{conclusion}{\section*{Conclusion}\normalsize\sf}{}
\newcommand{\References}[1]{\begin{flushleft}{\large References\\}\vspace*{2mm}\small #1 \end{flushleft}}
\newcommand{\chapterDSSN}[2]{\chapter[\sf\normalsize #1\\ \footnotesize \hspace*{5mm}by #2 \sf\normalsize][]{#1\\}\rhead[\fancyplain{}{\sf\footnotesize \center{#1}}]{\fancyplain{}{\sffamily\thepage}}\lhead[\fancyplain{\kopf}{\sffamily\thepage}]{\fancyplain{\kopf}{\sf\footnotesize \center{#2}}}}

\newcommand{\figureDSSN}[5]{\begin{figure}[#4]
\centering
\includegraphics*[#5]{#1}
\caption{#2}
\label{#3}
\end{figure}}

\renewcommand{\chaptername}{}
\newcommand{\acknowledgments}[1]{\vspace*{5mm}\noindent\begin{bf}Acknowledgments. \end{bf} #1}

\begin{document}
\sf

\chapterDSSN{A continuous population of variable stars up to about 1.5 mag 
above the horizontal branch?}{L. Baldacci, G. Clementini, E.V. Held, 
M. Marconi, L. Rizzi}

\Authors{L. Baldacci$^1$, G. Clementini$^2$, E.V. Held$^3$, M. Marconi$^4$
L. Rizzi$^3$} 
\Address{$^1$ Bologna University, via Ranzani 1, I-40127 Bologna\\
$^2$ INAF - Bologna Observatory, via Ranzani 1 I-40127 Bologna\\
$^3$ INAF - Padova Observatory, vicolo dell'Osservatorio 5 I-35122 Padova\\
$^4$ INAF - Capodimonte Observatory, via Moiariello 16 I-80131 Napoli}

\noindent
\begin{abstract}
Increasing samples of pulsating variable stars populating the classical 
instability strip from the horizontal branch to a few magnitudes brighter are 
being found in several Local Group galaxies, irrespective of the 
galaxy morphological type. We will review the observational 
scenario focusing in particular on the Anomalous Cepheids and related objects.
\end{abstract}

\noindent
\begin{introduction}
In recent years many Local Group (LG) galaxies have been surveyed looking 
for variables stars, however the observational scenario is rather 
inhomogeneous. In fact, while several of the dwarf Spheroidal (dSph) 
companions of the Milky Way had been studied in the early eighties, the advent 
of CCD detectors and wide field cameras, and the development of new
powerful methods for the detection of the variables (e.g. the image 
subtraction techniques), have prompted for 
new surveys whose results, however, often are not yet published. 
On the other side, the study of the variable stars in the Irregular 
galaxies was impossible until a few years ago, except in the 
Magellanic Clouds. Thus recent and good quality photometric data 
exist only for a few of these galaxies. Moreover, the
Magellanic Clouds benefited from the systematic 
surveys of the microlensing studies (e.g. MACHO:Alcock et al. 1996; OGLE:
Udalsky et al. 1997) thus samples for these two galaxies are much more well 
studied and complete than for others.

A number of different types of pulsating variables lie in the portion of the 
HR diagram brighter than the RR Lyr stars. We will briefly describe 
here the different types.

Anomalous Cepheids (ACs) are metal poor ($Z\sim 10^{-4}$) helium burning 
stars about 1 mag brighter than the RR Lyr stars, showing a range in period
from 0.3 d to 2 d. First observed in the Draco dSph by Baade \& Swope (1961),
they occur in all the known dSphs (Pritzl et al. 2002, and references therein). 
On the contrary they are remarkably rare in Globular Clusters (GCs) where 
only one confirmed AC has been found in the low density globular cluster 
\ngc5466 (Zinn \& Dahn 1976). 
Zinn \& Searle (1976) have named these variables Anomalous Cepheids because they 
fail to follow the period-luminosity (P/L) relation of Classical and 
Population II 
Cepheids (P2C). Because ACs do no show a different morphology of the light 
curves whether they pulsate in 
the fundamental or in the first overtone mode, the P/L 
relations (Nemec et al. 1988, Nemec et al .1994, 
Bono et al. 1997, Pritzl et al. 2002) are the only way to distinguish 
their pulsation mode, that, however, still remains uncertain for many of them.  
Theory and observations suggest that ACs are 2 or 3 times 
more massive than the RR Lyr stars, but we still lack of precise estimates
(Nemec et al. 1988 and reference therein).
Masses could allow us to disentangle between the two possible 
scenarios proposed for the origin of ACs: they are 
relatively young and massive stars ($\sim$ 1 Gyr, Norris \& 
Zinn 1975, M$<$ 2.5 M$\odot$) or they resulted from mass transfer in 
binary systems formed by old stars ($\sim 10$ Gyrs, Renzini et al. 1977). 
In the latter case 
their masses could not exceed twice the turn-off mass of the system 
($\sim 1.6 \rm{M_{\odot}}$ for Spheroidal galaxies, see Wallerstein \& Cox 
1984 for details). The question about the origin of the ACs in dSphs 
remains unsettled because none of these two
hypotheses could be definitely ruled out (Da Costa 1988).  

P2Cs are very common in Galactic GCs. They are very old, metal-poor, low mass 
(M$\leq$ 0.6 M$\odot$) stars that cross the instability strip while 
evolving from the blue tail of the HB towards the 
Asymptotic Giant Branch (see Wallerstein \& Cox, 1984). 
With the exception of Ursa Minor, dSphs generally 
do not have HBs extending to the blue enough to produce P2Cs, but  
they are found in the dSphs that host globular clusters, namely 
Fornax and Sagittarius.

Two further types of pulsating variables lie in the portion of the HR
diagram brighter than the HB of the LG dwarf Irregular (dIrr) galaxies, 
they are the short period Cepheids (SPCs) and the low luminosity (LL) 
Cepheids. Smith et al. (1992) defined SPCs a conspicuous population of 
variables in the Small Magellanic Cloud  with periods between 
0.6 d to 10 d, that did not follow the ACs P/L relations and fell instead 
on the extension to short periods of the Classical Cepheids P/L relations.  
SPCs have been found in many dIrrs since, they extend to longer periods  
and are brighter than the ACs. Both ACs and SPCs are helium burning stars,
but the former have experienced the helium flash (namely have masses 
$M \leq 2.5 \rm{M_{\odot}}$), while 
the latter are blue loop stars that have ignited helium in non degenerate 
conditions ($M \geq 2.5 \rm{M_{\odot}}$). SPCs should represent 
the low mass tail of the Classical Cepheids in low metallicity 
systems (Gallart et al. 1999, Dolphin et al. 2002).
 
LL Cepheids were observed for the first time in the dIrr galaxy NGC6822 by 
Clementini et al.(2003a). They have small amplitudes (0.1 - 0.4 mag in V),  
are fainter and have shorter periods than the SPCs: they seem to merge in 
magnitude with the RR Lyr stars 
forming a continuum in the classical instability strip. 
Due to their small amplitudes, LL Cepheids are difficult to detect, thus 
they may have been missed in dIrr galaxies where crowding is severe.
In the P/L plane the LL Cepheids occupy 
the short-period low-luminosity region where ACs and Classic Cepheids
P/L relations merge (see Fig.2) and where also fall the 4 ACs detected in the 
LMC by Clementini et al. (2003b), and the SPCs recently found in Phoenix by
Gallart et al. (2003).
Because of their low luminosities LL Cepheids could be the counterpart of 
ACs in dIrrs, indeed models for ACs predict variables with such 
small amplitudes and low luminosities (Fiorentino et al. 2003). 
\end{introduction}

\begin{spheroidal} 
\figureDSSN{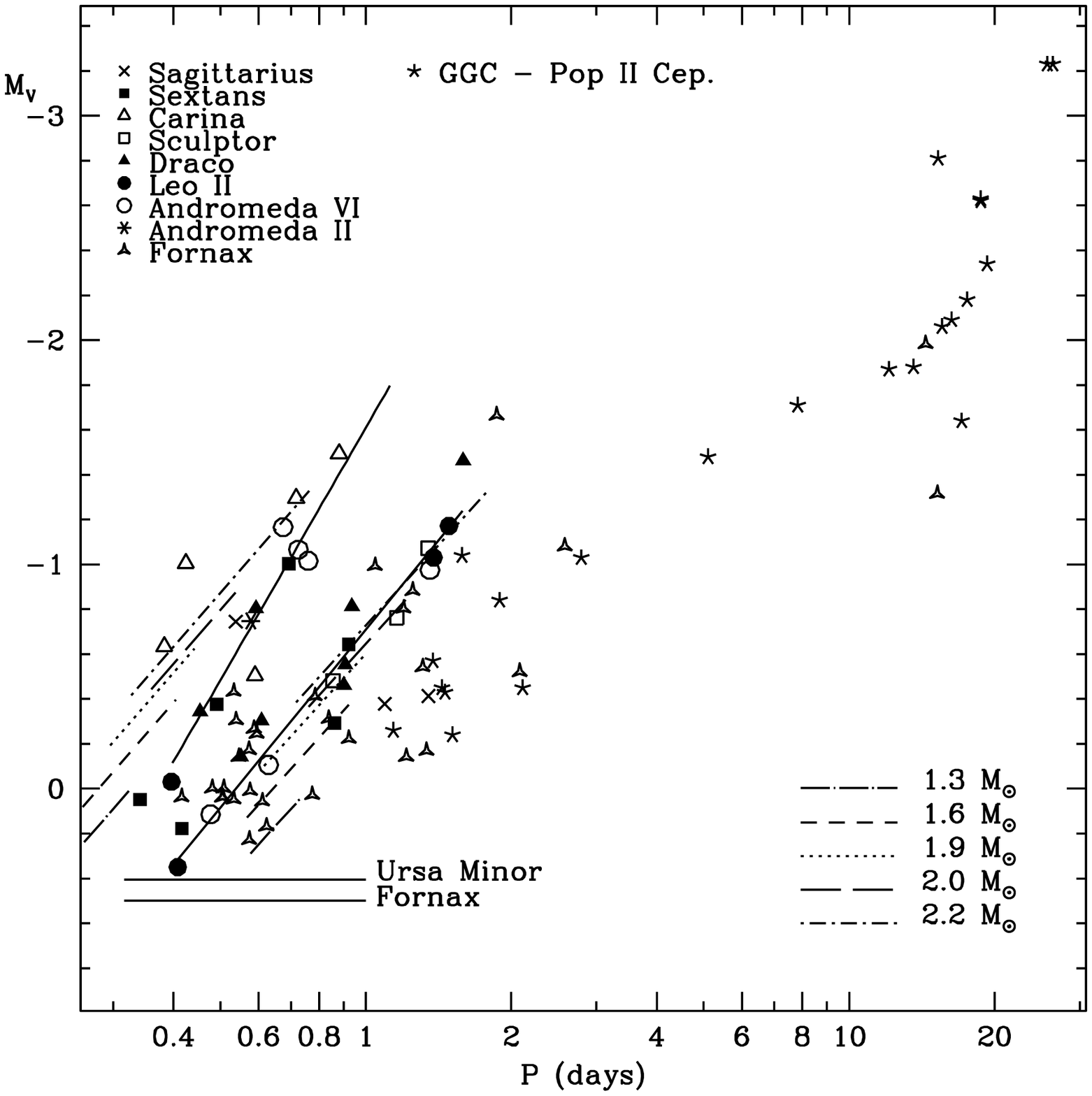}{P/L distribution of the ACs in various dSphs.
The slanted solid lines are the ACs P/L relations by
Pritzl et al. (2002), the horizontal solid lines mark the RR Lyr average 
luminosity in Ursa Minor and Fornax. Theoretical   
boundaries of the ACs instability strips are from Marconi et al. (2003, in 
preparation).}{fig:sp}{!ht}{bb=40 166 568 694, width=100mm}
All dwarf Spheroidals studied so far are found to host ACs, although not 
in very large numbers (And VI: Pritzl et al. 2002; Leo II: 
Siegel \& Majewski 2000; Draco: Zinn \& Searle 
1976, Kinemuchi et al. 2002; Ursa Minor: Nemec et al 1988; Carina: Saha et 
al. 1986, Dall'Ora et al. 2003; Sculptor: Swope 1968, Kaluzny et al. 1995;
Sextans: Mateo et al. 1995a, Sagittarius: Mateo et al. 1995b, Layden \& 
Sarajedini 2000; AndIII: Da Costa et al. 2002; And II: Pritzl et al. 2003;
Leo I: Hodge \& Wright 1978; Fornax: Light et al. 1986, Bersier \& Wood 2002, 
Clementini et al. 2003c, Mackey \& Gilmore 2003).
Figure \ref{fig:sp} shows the P/L relation of the ACs in the V band, 
drawn from the above literature data generally selecting only variables
with accurate data.
We have adopted the distance scale in Pritzl et al. (2002) which is 
consistent with a distance modulus of 18.5 for the LMC.
The distance scale has a fundament role 
both in the slope and zero point of the ACs P/L relations (slanted solid lines
in Figure \ref{fig:sp}), 
and in the pulsation mode determination. The horizontal lines mark the mean 
level of RR Lyr stars in the metal poorer (Ursa Minor) and the metal richer 
(Fornax) galaxies and show that there is no clear separation between the 
distributions of ACs and HB variables. Also shown in the figure are the 
boundaries of the theoretical instability strip of ACs models  
with different masses computed by Marconi et al. (2003, in prep.).  
The Leo\,I ACs are not displayed and are discussed more in detail in the 
next section. The Galactic GCs P2Cs collected by Nemec et al. (1994) 
are also shown in figure \ref{fig:sp}. 
Fornax and Sagittarius are so far the only dSphs known to host 
GCs, and found to contain both ACs and P2Cs. Two of the variables
in Sagittarius are in the   
cluster M54 (Layden \& Sarajedini 2000). They had been originally classified 
as a candidate AC and a candidate P2C, but
clearly lie both in the P2Cs region. 
Fornax variables present a more complex scenario since many of the field
variables classified as ACs by Bersier \& Wood (2000) lie instead on the
P/L relation of the P2Cs. 
If confirmed when more accurate photometric data will become available, this 
would thus be the first identification of P2Cs in the field of a dSph.

$Leo I$\\ 
Leo I has a dominant young and intermediate-age stellar population.
The galaxy has a bulk of candidate variables 2.4 mag 
brighter than the HB, and about 1 mag brighter than the ACs found in the 
other dSphs (Lee et al. 1993, Baldacci et al. 2003). 
Caputo et al. (1999) demonstrate that they are still consistent     
with ACs models, but Gallart et al. (1999) claim that they are
brighter enough to be SPCs. 
The published light curves for these variables (Hodge \& Wright 1978) 
are affected by a large scatter, thus 
the presence of variables more massive than the ACs in Leo I is a
hypothesis that needs to be confirmed on the basis of more accurate
data (Clementini et al. 2004, in preparation).  
\end{spheroidal}

\begin{irregular}
\figureDSSN{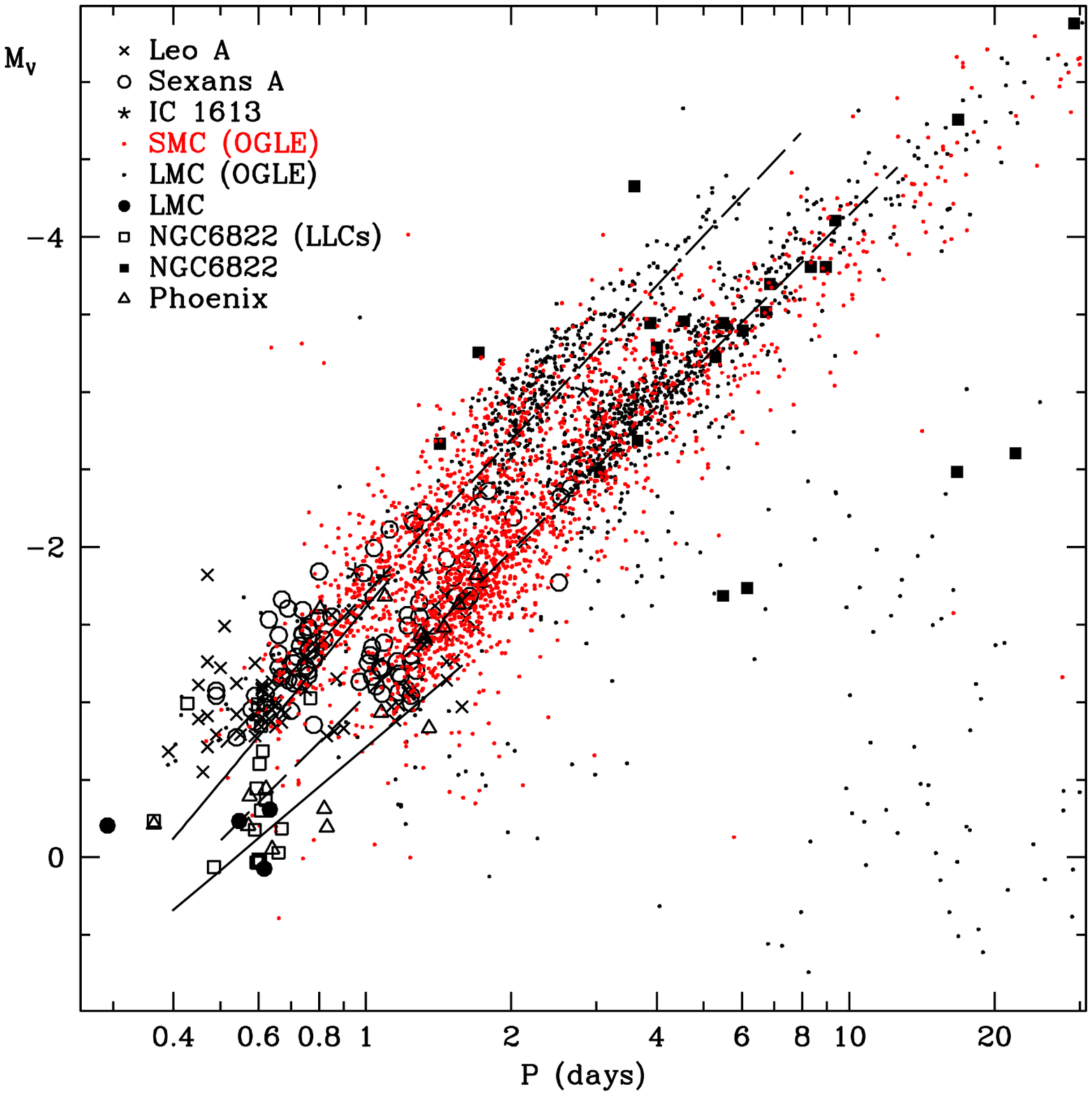}{P/L distribution of Cepheids with period shorter than 
20 d found in LG Irregular galaxies.}{fig:ir}{!t}{bb=40 166 568 694, width=100mm}
Figure \ref{fig:ir} shows the P/L distribution of Cepheids with period 
shorter than 20 d found in the LG Irregular galaxies 
(IC1613: Dolphin et 
al. 2001; Leo A: Dolphin et al. 2002; Sextans A: Dolphin et al. 2003; 
SMC: Udalski et al. 1999; LMC: Udalski et al. 1999, Clementini et al. 2003b; 
NGC6822: Antonello et al. 2002, Clementini et al. 2003a; Phoenix: Gallart et 
al. 2003). The dashed-dotted lines are the P/L relations 
followed by the SPCs 
with period shorter than 2 days in the SMC (see Dolphin et al. 2003 for 
detail). We have extended them to larger periods (up to $\sim$ 10 d)
since they very well represent the general distributions shown 
in Figure \ref{fig:ir}. The solid lines are the 
P/L relations for ACs in dSphs by Pritzl et al. (2002). The P/L relations
of the first overtone ACs and Cepheids are similar to each other, the 
fundamental mode relations show instead some 
differences. The LL Cepheids in NGC6822 appear to be consistent 
both with the ACs and SPCs relations.               
\end{irregular}

\begin{conclusion}
The classical instability strip appears to be continuously populated in the 
dIrrs with the LL Cepheids filling the gap between RR Lyr stars and SPCs.
An instability strip uniformly populated as observed in NGC 6822 
or in Phoenix probably reflects the continuous star formation process
occurring in these galaxies. On the 
other side, ACs with luminosities close to the RR Lyr stars have been found in 
some dSphs but they are only a few. Thus they may represent the  
the tail at low masses of the ACs mass distribution, more than 
a continuity in the star formation process.      
The actual nature of the LL Cepheids still remains unclear since 
they cannot be distinguished from ACs and SPCs on the basis of the P/L
relations.
\end{conclusion}

\acknowledgments{
We are grateful to C. Gallart for providing us the data for the SPCs in Phoenix,
and to D. Bersier for helpful discussions on the
ACs and P2Cs in Fornax.
}
\References{
Alcock, C. et al. 1996, AJ 111, 1146 \\
Antonello, E., Fugazza, D., Mantegazza, L., Stefanon, M., Covino, S. 2002, A\&A 386, 860 \\
Baade, W., Swope, H.H. 1961, AJ 66, 300\\
Baldacci, L. et al. 2003, in ``Stars in Galaxies'', astro-ph/0305506 \\
Bersier, D., Wood, P.R. 2002, AJ 123, 840\\
Bono, G., Caputo, F., Santolamazza, P., Cassisi, S., Piersimoni, A. 1997, AJ 113, 2209 \\
Caputo, F., Cassisi, S., Castellani, M., Marconi, G., Santolamazza, P. 1999, AJ 117, 2199\\ 
Clementini, G., Held, E.V., Baldacci, L., Rizzi, L. 2003a, ApJ 588L, 85 \\ 
Clementini, G. et al. 2003b, AJ 125, 1309 \\
Clementini, G. et al. 2003c, in ``Variable stars in the Local Group'', astro-ph/0310545 \\
Da Costa, G.S. 1988, IAUS 126, 217 \\
Da Costa, G.S., Armandroff, T.E., Caldwell, N. 2002, AJ 124, 332\\
Dall'Ora, M. et al. 2003, AJ 126, 197 \\
Dolphin, A.E. et al. 2001, ApJ 550, 554 \\
Dolphin, A.E. et al. 2002, AJ 123, 3154 \\
Dolphin, A.E. et al. 2003, AJ 125, 1261 \\
Fiorentino, G., Caputo, F., Marconi, M. 2003, in ``Stars in Galaxies'', in press \\
Gallart, C., Freedman, W.L., Aparicio, A., Bertelli, G., Chiosi, C., 1999, AJ 118, 2245 \\ 
Gallart, C. et al. 2003, in ``Variable stars in the Local Group'', in press\\
Hodge, P.W., Wright, F.W. 1978, AJ 83, 228 \\
Kaluzny, J. et al. 1995, A\&AS 112, 407 \\
Kinemuchi, K. et al. 2002, ASP Conf. Ser. 259, 130 \\
Layden, A.C., Sarajedini, A., 2000, AJ 119, 1760  \\
Lee, M.G. et al. 1993, AJ 106, 1420 \\
Light, R.M., Armandroff, T.E., Zinn, R. 1986, AJ 92, 43\\
Mackey, A.D., Gilmore, G.F., 2003 astro-ph/0307275 \\
Mateo, M., Fischer, P., Krzeminski, W. 1995a, AJ 110, 2166\\
Mateo, M. et al. 1995b, AJ 110, 1141 \\
Nemec, J.M., Wehlau, A., Mendes de Oliveira, C. 1988, AJ 96, 528\\
Nemec, J.M., Nemec, A.F.L., Lutz, T.E. 1994, AJ 108, 222 \\
Norris, J.,  Zinn, R. 1975, ApJ 202, 335 \\
Pritzl, B.J., Armandroff, T.E., Jacoby, G.H., Da Costa, G.S. 2002, AJ 124, 1464\\ 
Pritzl, B.J., Armandroff, T.E., Jacoby, G.H., Da Costa, G.S. 2003, astro-ph 0310620\\
Renzini, A., Mengel, J.G., Sweigart, A.V. 1977, A\&A 56, 369\\
Saha, A., Monet, D.G., Seitzer, P. 1986, AJ 92, 302 \\
Siegel, M.H., Majewski, S.R 2000, AJ 120, 284 \\
Swope, H.H. 1968, AJS 73, 204 \\ 
Smith, H.A., Silbermann, N.A., Baird, S.R., Graham, J.A. 1992, AJ 104, 1430\\
Udalsky, A., Kubiak, M., Szymanski, M. 1997, AcA 47, 319\\
Udalski, A. et al. 1999, AcA 49, 201\\
Wallerstein, G., Cox, A.N. 1984, PASP 96, 677 \\
Zinn, R., Dahn, C.C. 1976, AJ 81,527\\
Zinn, R., Searle, L. 1976, ApJ 209, 734\\
}

\end{document}